\documentstyle[aps,preprint,floats,prc]{revtex}

\input{boxedeps}
\HideDisplacementBoxes
\SetRokickiEPSFSpecial

\begin{document}

\draft
\preprint{\vbox{\it \null\hfill\rm
DOE/ER/40561-271-INT96-00-132}\\\\}

\title{Breakup reactions of the halo nuclei $^{11}$Be and $^{8}$B}

\author{
Kai Hencken
\and 
George Bertsch}
\address{
Institute for Nuclear Theory,
University of Washington,
Seattle, WA 98195, USA
}
\author{
Henning Esbensen}
\address{
Physics Division,
Argonne National Laboratory,
Argonne, IL 60439, USA
}

\date{\today}

\maketitle

\begin{abstract}
We calculate the nuclear induced breakup of $^{11}$Be and $^{8}$B
using a more realistic treatment of the diffraction and stripping
processes than in previous work.  The breakup is treated in the
eikonal approximation with a profile function calculated from a
realistic optical potential at low energies and from free
nucleon-nucleon cross sections at high energies. This treatment gives
a good description of measured breakup cross sections, as well as the
longitudinal momentum distribution of the core-like fragments, which
is narrower than predicted in the transparent limit.  The real part
of the potential is found to be significant and enhances the
diffractive breakup at low energies.
\end{abstract}

\narrowtext
\tighten
\section{Introduction}
\label{Sec:intro}

Breakup reactions of light nuclei near the neutron or proton drip line
have demonstrated the existence of nuclei with a halo composed of one
or two loosely bound nucleons. The momentum distribution of the
fragments becomes much narrower in halo nuclei, and it is common to
interpret these distributions in terms of the momentum-space wave
function of the halo nucleus.  This picture is overly simplified and
more microscopic modeling of the reaction process is needed to
interpret the distributions \cite{esbensen96a,esbensen96b}.

In this work we investigate this question further using realistic
projectile-target interactions.  We calculate the breakup of
$^{11}$Be and $^{8}$B by a nuclear target using the eikonal
approximation for the reaction theory. This is essentially the model
developed by Serber \cite{serber47} and Glauber \cite{glauber55} to
describe the breakup of the deuteron, improved beyond the black-disk
approximation used in that original work \cite{sitenko90}. Our
interaction will be taken from a realistic optical potential at low
energies and from total nucleon-nucleon cross sections at high
energies.

There are two different processes by which the breakup occurs, namely
stripping and diffraction. We present the theory of these processes
in the next section. Stripping depends only on the absorption
interaction, while diffraction is quite sensible to the real part as
well.  Our interaction includes both, as described in
Sec.~\ref{Sec:potential}.  Details of the numerical calculations are
described in Sec.~\ref{Sec:calculation}. Results for the total cross
section are discussed in Sec.~\ref{Sec:total}, the longitudinal
momentum distributions in Sec.~\ref{Sec:Be11} and~\ref{Sec:B8}.

\section{Probabilities and Cross Sections in the Diffractive Model}
\label{Sec:serber}

We consider a single-particle model of the halo nucleus.  The ground
state is described by a wave function $\phi_0(\vec{r})$ which depends
on the relative coordinate $\vec{r}$ between the nucleon and the
core.

After interacting with a target, the wave function of the halo nucleus
in its rest frame has the form,
\begin{equation}
\Psi(\vec{r},\vec{R}) = S_n(\vec{b}_n) S_c(\vec{b}_c) \phi_0(\vec{r}) 
\ ,
\label{Eq:final}
\end{equation}
where $\vec{R}$ is the coordinate of the center of mass of the halo
nucleus, and $\vec{b}_c$ and $\vec{b}_n$ are the impact parameters of
the core and the nucleon with respect to the target nucleus, i.~e.
$\vec{b}_n=\vec{R}_\perp + \vec{r}_\perp A_c/(A_c+1)$ and
$\vec{b}_c=\vec{R}_\perp - \vec{r}_\perp /(A_c+1)$, where $A_c$ is the
mass number of the core.  The two profile functions, $S_n(\vec{b}_n)$
for the nucleon and $S_c(\vec{b}_c)$ for the core, are generated by
interactions with the target nucleus.  In the eikonal approximation,
they are defined by the longitudinal integrals over the corresponding
potentials:
\begin{equation}
S(\vec{b}) = \exp \left[\frac{-i}{\hbar v}
\int dz V(\vec{b} + z  \hat{z})  \right] \ ,
\label{Eq:profile}
\end{equation}
where $v$ is the beam velocity. The potential $V$ is the full optical
potential, including the Coulomb potential and the real and imaginary
parts of the nuclear potential. The scattering wave function is the
difference between eq.~(\ref{Eq:final}) and the wave function of the
undisturbed beam,
\begin{equation}
\Psi_{scat} = (S_n S_c -1 ) \phi_0 \ .
\end{equation}
Elastic and diffractive scattering are calculated by taking overlaps
of $\Psi_{scat}$ with different final states.  For elastic scattering,
we take the overlap with the halo nucleus in its ground state, but
with some arbitrary transverse momentum $\vec{K}_\perp$.  This is
\begin{equation}
a(\vec{K}_\perp) = 
\int d^2 \vec{R}_\perp e^{-i\vec{K}_\perp\cdot \vec{R}}
\int d^3\vec{r} \phi_0^*(\vec{r}) (S_c S_n-1) \phi_0(\vec{r})\ .
\end{equation}
The differential and total elastic cross sections are then given by
\begin{equation}
{d\sigma_{\text{el.}}\over d^2\vec{K}_\perp} =  
\frac{|a(\vec{K}_\perp)|^2}{(2\pi)^3} \ , 
\end{equation}
\begin{equation}
\sigma_{\text{el.}} = \int d^2\vec{R}_\perp \left|
\int d^3\vec{r} \phi_0^*(\vec{r}) (S_c S_n - 1) \phi_0(\vec{r})
\right|^2 \ .
\end{equation}
In the case of $^{8}$B, where the proton is in a $p$ state, we have
to sum also over the final and average over the initial $M$ states:
\begin{eqnarray}
&&\sigma_{\text{el.}} = \frac{1}{2 L_0+1} \sum_{M_0,M_0'} \nonumber\\
&&\int d^2\vec{R}_\perp \left|
\int d^3\vec{r} \phi_{0,M_0'}^*(\vec{r}) (S_c S_n - 1) 
\phi_{0,M_0}(\vec{r}) \right|^2 \ .
\end{eqnarray}

For diffractive breakup the final state depends on the relative
momentum $\vec{k}$ of nucleon and core in their center-of-mass frame
as well as on the transverse momentum $\vec{K}_\perp$ of the center
of mass.  Writing the continuum nucleon-core wave function as
$\phi_{\vec{k}}(\vec{r})$, the diffractive breakup cross section is
given by
\begin{eqnarray}
&&\frac{d\sigma_{\text{diff.}}}{\left(d^2\vec{K}_{\perp} d^3\vec{k}
\right)}= \frac{1}{(2\pi)^5} \frac{1}{2 L_0+1} \sum_{M_0}
\nonumber\\
&&\left|
\int d^3\vec{r} d^2\vec{R}_{\perp} 
e^{-i\vec{K}_\perp\cdot\vec{R}_\perp} 
\phi_{\vec{k}}^*(\vec{r}) S_c S_n \phi_{0,M_0}(\vec{r}) \right|^2 \ .
\end{eqnarray}
Here the continuum wave function is normalized asymptotically to a
plane wave, $ \phi_{\vec{k}}\sim\exp(i\vec{k}\cdot\vec{r}) $.  If we
are only interested in the relative momentum distribution, i.e.  in
$\vec{k}$, we can integrate over $\vec{K}_{\perp}$ to get
\begin{eqnarray}
\frac{d\sigma_{\text{diff.}}}{d^3\vec{k}} 
&=& \frac{1}{(2\pi)^3} \frac{1}{2 L_0+1} \sum_{M_0} \nonumber\\
&&\int d^2\vec{R}_\perp \left|
\int d^3\vec{r} \phi_{\vec{k}}^*(\vec{r}) S_c S_n \phi_{0,M_0}
(\vec{r}) \right|^2\ .
\label{Eq:DDiff}
\end{eqnarray}
A convenient expression for the total diffractive cross section can
be derived if $\phi_0$ is the only bound state of the system.  This
is
\begin{eqnarray}
\sigma_{\text{diff.}} &=& \frac{1}{2 L_0+1} \sum_{M_0}
\int d^2\vec{R}_\perp \nonumber\\
&&\Bigg[\int d^3\vec{r} \phi_{0,M_0}(\vec{r})^* 
\left|S_c S_n\right|^2 \phi_{0,M_0}(\vec{r}) \nonumber\\
&&- \sum_{M_0'}
\left|\int d^3\vec{r} \phi_{0,M_0'}(\vec{r})^* S_c S_n 
\phi_{0,M_0}(\vec{r}) \right|^2 \Bigg] \ .
\label{Eq:diff}
\end{eqnarray}

Other contributions to the total cross section come from absorption,
present when the eikonal factors have moduli less than 1.  There are
three of these so-called stripping processes.  The nucleon-absorption
cross section, differential in the momentum of the core, is given by
\begin{eqnarray}
\frac{d\sigma_{\text{n-str.}}}{d^3\vec{k}_c} 
&=& \frac{1}{(2\pi)^3} \frac{1}{2 L_0+1} \sum_{M_0} \int d^2\vec{b}_n 
\left[1- \left|S_n(\vec{b}_n)\right|^2 \right] \times\nonumber\\
&&\left| \int d^3\vec{r} e^{-i \vec{k}_c\cdot\vec{r}} 
S_c(\vec{b}_c) \phi_{0,M_0}(\vec{r}) \right|^2 \ .
\label{Eq:strip}
\end{eqnarray}
The corresponding total cross section for stripping of the nucleon is 
\begin{eqnarray}
\sigma_{\text{n-str.}} &=& \frac{1}{2 L_0+1} \sum_{M_0}
\int d^2\vec{b}_n
\left[1- \left|S_n(\vec{b}_n) \right|^2 \right] \times\nonumber\\
&&\int d^3\vec{r} \phi_{0,M_0}(\vec{r})^* 
\left|S_c(\vec{b}_c)\right|^2 \phi_{0,M_0}(\vec{r}) \ .
\end{eqnarray}
The stripping of the core is expressed in a similar way, interchanging
subscripts $n$ and $c$.

Finally, the expression for absorption of both nucleon and core is
given by
\begin{eqnarray}
\sigma_{\text{abs.}} &=& \frac{1}{2 L_0+1} \sum_{M_0}
\int d^2\vec{b}_c 
\left[ 1 - \left| S_c(\vec{b}_c) \right |^2\right] \times \nonumber\\
&&\int d^3\vec{r} \phi_{0,M_0}(\vec{r})^* 
\left[ 1 - \left| S_n(\vec{b}_n) \right |^2\right] 
\phi_{0,M_0}(\vec{r}) \ .
\end{eqnarray}

All of the above contributions have been written in the form of an
integration over some transverse coordinate, so the integrand may be
interpreted as an impact-parameter-dependent probability. Note,
however, that for the diffraction and the elastic scattering, the
integration is over the impact parameter with respect to the center
of mass, whereas it is over the impact parameter with respect to the
nucleon or the core for the stripping and absorption processes. For
the total cross sections we can change the integration variable to
the center-of-mass impact parameter in these cases. All the different
probabilities together with the one for no interaction then add up to
one, showing that all possible processes are included in this scheme.

Note also that the diffractive differential cross section
(eq.~(\ref{Eq:DDiff})) is expressed as a function of the relative
momentum, i.~e. the momentum of the nucleon or the core in the
nucleon-core center-of-mass frame.  No expression of the same form
exists for the nucleon or core momenta due to the interplay with the
diffraction of the center-of-mass motion.

\section{The potential for the Nucleon-Target and Core-Target 
Interaction}
\label{Sec:potential}

Evaluation of the profile functions requires a potential model for the
interaction between the target nucleus and the constituents of the
halo nucleus.  At low energies, extending up to about 100 MeV/$n$, one
can find optical potentials that are fit to nucleon-nucleus
scattering.  We use the potential of ref.~\cite{ch89}, which was fit
to scattering data in the range of 10 to 60 MeV.  The potential has
the usual Woods-Saxon form, with volume and surface imaginary terms.
We drop the spin-orbit term, which has a small effect on
spin-independent observables.  This potential can be used as it stands
for the target-nucleon interaction.  We apply it to the core-target
interaction by folding it with the core density distribution,
\begin{eqnarray}
V_c(r) = \int d^3\vec{x} \rho_c(x) 
 V_{\text{op}}(|\vec{r} - \vec{x}|) \ .
\end{eqnarray}
For the core density we use a harmonic oscillator density with
parameters taken from the charge distribution of the core nucleus
\cite{devries87} ($a$=2.5~fm and $\alpha$=0.61~fm for the $^{10}$Be
and similar for $^7$Be).

At high energies, many-body effects on the effective interaction are
small, and we may use the free nucleon-nucleon interaction to
generate the potentials.  Following
ref.~\cite{bertsch90,esbensen92,bertsch89}, we ignore the finite
range of the NN interaction and take the potential to be proportional
to the density of the target
\begin{equation}
V_\rho(r) = V_0 \rho_t(r) \ .
\end{equation}
We have either a harmonic oscillator or a Fermi form factor for the
target density distribution and use the tabulated width and
diffuseness \cite{devries87}. We assume that protons and neutrons
have the same density distribution.

The imaginary part of $V_0$ can be determined from free
NN cross sections. The result for the neutron-target
interaction is
\begin{equation}
\text{Im}\left(V_n(r)\right) = - \frac{1}{2} \hbar v 
\Bigl(Z_T \sigma_{np} + N_T \sigma_{nn}\Bigr)  \rho_t(r) \ ,
\end{equation}
where the spatial integral of the target density is normalized to
one, and a similar formula for the proton-target interaction.  Here
$Z_T$ and $N_T$ are the proton and neutron numbers, respectively, of
the target nucleus, and the $\sigma$'s are the total NN cross
sections.  The core-nucleus interaction is generated by folding as it
was done earlier for the optical potential.  The total NN cross
sections are calculated from the velocity-dependent parameterization
given in ref.~\cite{giacomelli70}.

The real part of the potential becomes small above 200 MeV and is
negligible at 800 MeV \cite{ray79}. We have neglected it in our
calculations of the breakup at high energy. We shall also neglect the
Coulomb part of the interaction, which makes a negligible
contribution to breakup cross sections on light
nuclei. Unfortunately, the Coulomb field may nevertheless affect the
transverse momentum distributions. We therefore restrict our study to
total cross sections and longitudinal momentum distributions.

\section{Numerical calculation of the probabilities}
\label{Sec:calculation}

As seen in Sec.~\ref{Sec:serber}, the cross sections all require
calculating probability functions depending on the impact parameter,
that have the form
\begin{equation}
\frac{d^3P_{\Omega}(b)}{d^3\vec{k}} = 
\frac{1}{(2\pi)^3} \left| 
\int d^3\vec{r}\  \phi_{\vec{k}}^*(\vec{r}) 
\Omega(\vec{b},\vec{r}_\perp) \phi_0(\vec{r})
\right|^2 \ ,
\label{Eq:pomega}
\end{equation}
with an appropriate chosen $\Omega$.

The wave functions $\phi_0$ and $\phi_{\vec{k}}$ are determined as
eigenfunction for a Woods-Saxon potential together with the Coulomb
potential of a homogeneous charged sphere in the case of $^{8}$B.

The ground state of
$^{11}$Be has spin-parity of $1/2^+$, so we take an s-wave,
\begin{equation}
\phi_0(r) = \frac{g(r)}{r} \ Y_{00} \ .
\end{equation}
With our geometry of the Woods-Saxon potential, ($R=2.7$ fm and
$a=0.52$ fm \cite{esbensen96a}), the empirical neutron binding energy
of 0.503 MeV is reproduces with a well depth of about $V_0=-61.1$ MeV.

For calculating diffractive cross sections, we expand the continuum
wave function $\phi_{\vec{k}}$ in the usual partial wave
representation,
\begin{equation}
\phi_{\vec{k}}(\vec{r}) = \frac{4\pi}{k} \sum_{L,M} i^L e^{-i\delta_L}
\frac{u_L(r)}{r} Y_{L,M}^*(\hat k) Y_{L,M}(\hat r) \ .
\end{equation}
Here the $\delta_L$ are the phase shifts and $u_L(r)$ are the (real)
solutions of the radial Hamilton operator. The probability is then
given by
\widetext
\begin{eqnarray}
\frac{d^3P_{\Omega}(b)}{d^3\vec{k}} &=& 
\frac{1}{(2\pi)^3 (2L_0+1)} \left(\frac{4 \pi}{k} \right)^2 
\times\nonumber\\
&&\sum_{M_0} \biggl| \int d^3\vec{r}  \sum_{L,M}
(-i)^L e^{i\delta_L} Y_{L,M}(\hat k)
Y_{L,M}^*(\hat r) \frac{u_L(r)}{r}
\Omega(\vec{b},\vec{r}_\perp) \frac{g(r)}{r} Y_{L_0,M_0}(\hat r)
\biggl|^2 \ .
\end{eqnarray}

As the direction of the impact parameter is not observed, we can
integrate over the transverse angle of the momenta $\varphi_k$ to get
the double-differential probability
\begin{eqnarray}
\frac{d^2P_{\Omega}(b)}{dk_l dk_\perp} &=& 
k_\perp \int d\varphi_k \frac{d^3P_{\Omega}(b)}{d^3\vec{k}}\\
&=& \frac{2 \pi k_\perp}{(2\pi)^3 (2L_0+1)} 
\left(\frac{4 \pi}{k} \right)^2 \times \nonumber\\
&&
\sum_{M,M_0} \biggl| \int d^3\vec{r} \sum_{L}
(-i)^L e^{i\delta_L} \tilde Y_{L,M}(\theta_k)
Y_{L,M}^*(\hat r) \frac{u_L(r)}{r}
\Omega(\vec{b},\vec{r}_\perp) \frac{g(r)}{r} Y_{L_0,M_0}(\hat r)
\biggr|^2.
\end{eqnarray}
In deriving this we have used the orthogonality of the spherical
harmonics with respect to integration over $\varphi_k$ and we define
$\tilde Y_{L,M}(\theta_k) \equiv Y_{L,M}(\hat k) e^{-iM\varphi_k}$,
i.e., the spherical harmonic without the $\varphi_k$-dependent part.

The numerical integration will be performed in spherical coordinates
as follows 
\begin{eqnarray}
&&\frac{d^2P_{\Omega}(b)}{dk_l dk_\perp} =
2 \pi k_\perp
\frac{1}{(2\pi)^3 (2L_0+1)} \left(\frac{4 \pi}{k} \right)^2 
\sum_{M,M_0} \biggl| 
\int dr \int d\cos\theta \times \nonumber\\
&&\left[\sum_L (-i)^L e^{i\delta_L} 
\tilde Y_{L,M}(\theta_k) u_L(r) g(r)
\tilde Y_{L,M}(\theta) \tilde Y_{L_0,M_0}(\theta) \right]
\int d\varphi \exp\left(i (M_0-M)\varphi\right)
\Omega(\vec{b},\vec{r}_\perp) \biggl|^2.
\end{eqnarray}
\narrowtext
As $\Omega$ is an even function in $\varphi$, we can replace
$\exp(i\varphi)$ by $\cos(\varphi)$ in the last expressions. Note
that the summation over $L$ and the integration over $\varphi$ are
independent of each other, and therefore the integration over
$\varphi$ has to be done only one time instead of $L$ times for the
sum.  There are several advantages to using this expression from a
numerical point of view. The number of $\varphi$-integrations is
lower by a factor of $L$ due to the factorization.  Only one
integration over the coordinates has to be done, which then has to be
squared. And finally, by doing the $\theta$-integration inside the
$r$-integration, we can minimize the number of wave-function
evaluations, which is normally a rather calculation-intensive
step. Angular momenta up to $l=5$ or $l=7$ have been used throughout
the calculation. The cross section for the different processes can
then be found by integrating over the impact parameter. Longitudinal
momentum distribution are found by integrating also over the
transverse momenta.  For these three integration we have used a
Gaussian integration with a fixed number of points.

For better numerical convergence in the calculation of diffractive
excitation, we replace the operator $\Omega=S_cS_n$, by
$\Omega=S_cS_n-1$, which are of course equivalent for transition
matrix elements.

In order to test the accuracy of our calculation we have compared the
results of the differential cross sections integrated over all $k$
and the separate calculation of the total cross section directly from
eq.~(\ref{Eq:diff}).  Both results were found to agree within a few
percent. Therefore our results should be accurate to a few percent.

For the calculation of the stripping probabilities, we use the same
eq.~(\ref{Eq:pomega}), replacing the scattering wave by a plane wave.
Thus the phase-shifts are set to zero and the continuum partial waves
$u_L(r)$ are replaced by the spherical Bessel functions, $j_L(kr)$.

We do the same in the case of $^{8}$B and use a $p$ wave for the
valence proton in the ground state. The parameters of the Wood-Saxon
potential in this case are $R=$2.48~fm and $a=$0.52~fm.  We include
also the Coulomb potential of a homogeneous charged sphere with the
same radius as the core in the calculation of bound and continuum
states.  The binding energy of 0.137~MeV is reproduced by a well depth
of about $V_0=-47.5$~MeV.

\section{Integrated cross sections}
\label{Sec:total}

\begin{figure}[t]
\begin{center}
~\ForceWidth{0.44\hsize}
\tBoxedEPSF{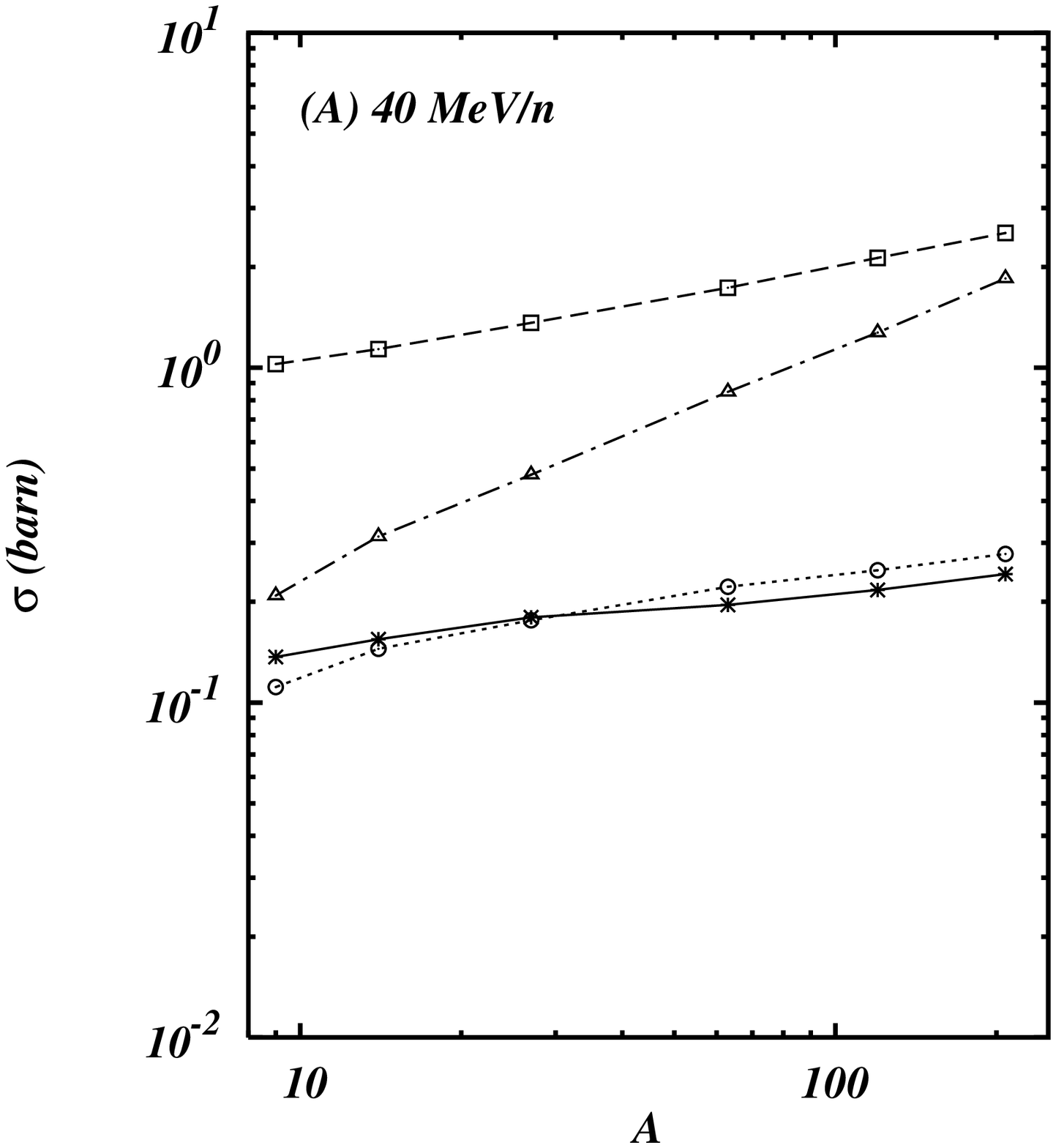}~~~~
\ForceWidth{0.44\hsize}
\tBoxedEPSF{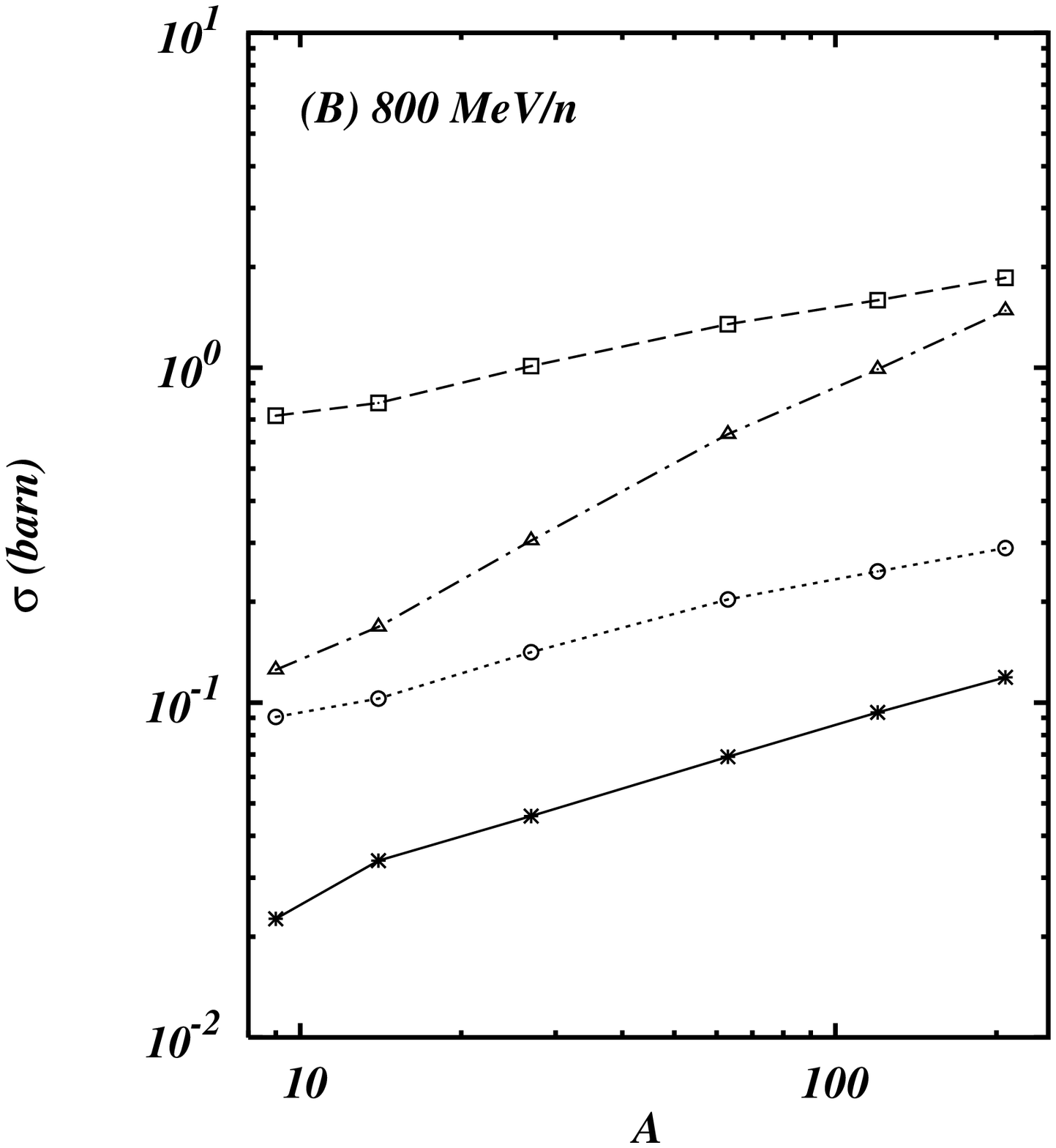}
\end{center}
\caption{Cross sections for the different nuclear induced breakup
processes as functions of the target mass number $A$. Results are
shown for a beam energy of 40 MeV/$n$ (A) and 800 MeV/$n$ (B). Shown
are the cross section for diffraction (solid line and stars), for the
stripping of the neutron (dotted line and circles) and the core
(dashed line and boxes) and for the absorption of both neutron and
core (dashed-dotted line and triangles).}
\label{Fig:adep}
\end{figure}

We first show our calculated results for the integrated breakup cross
sections for $^{11}$Be. The diffractive and the three stripping cross
sections are shown in Fig.~\ref{Fig:adep} as functions of the target
mass $A$.  The largest component is core stripping, as might be
expected.  The absorption of both core and neutron is next. The sum
of both vary with $A$ roughly as the geometric size, i.~e. as
$(A_c^{1/3} + A^{1/3})^2$. The absorption cross section increases
faster with $A$ at high energies than at low energies, showing the
deviation from a black disc.  The diffractive component is quite
small at 800 MeV/$n$, but it has the same size as the neutron
absorption at 40 MeV/$n$.  The difference is due to the real part of
the potential, which is important at low energies and increases the
diffractive scattering. At low energies diffraction and
neutron-stripping are of the same magnitude for all $A$.  This agrees
with other calculations \cite{yabana92,anne93} which find also that
both processes are almost the same up to energies of 100 MeV/$n$.
%
%

\begin{figure}[t]
\begin{center}
~\ForceWidth{0.44\hsize}
\tBoxedEPSF{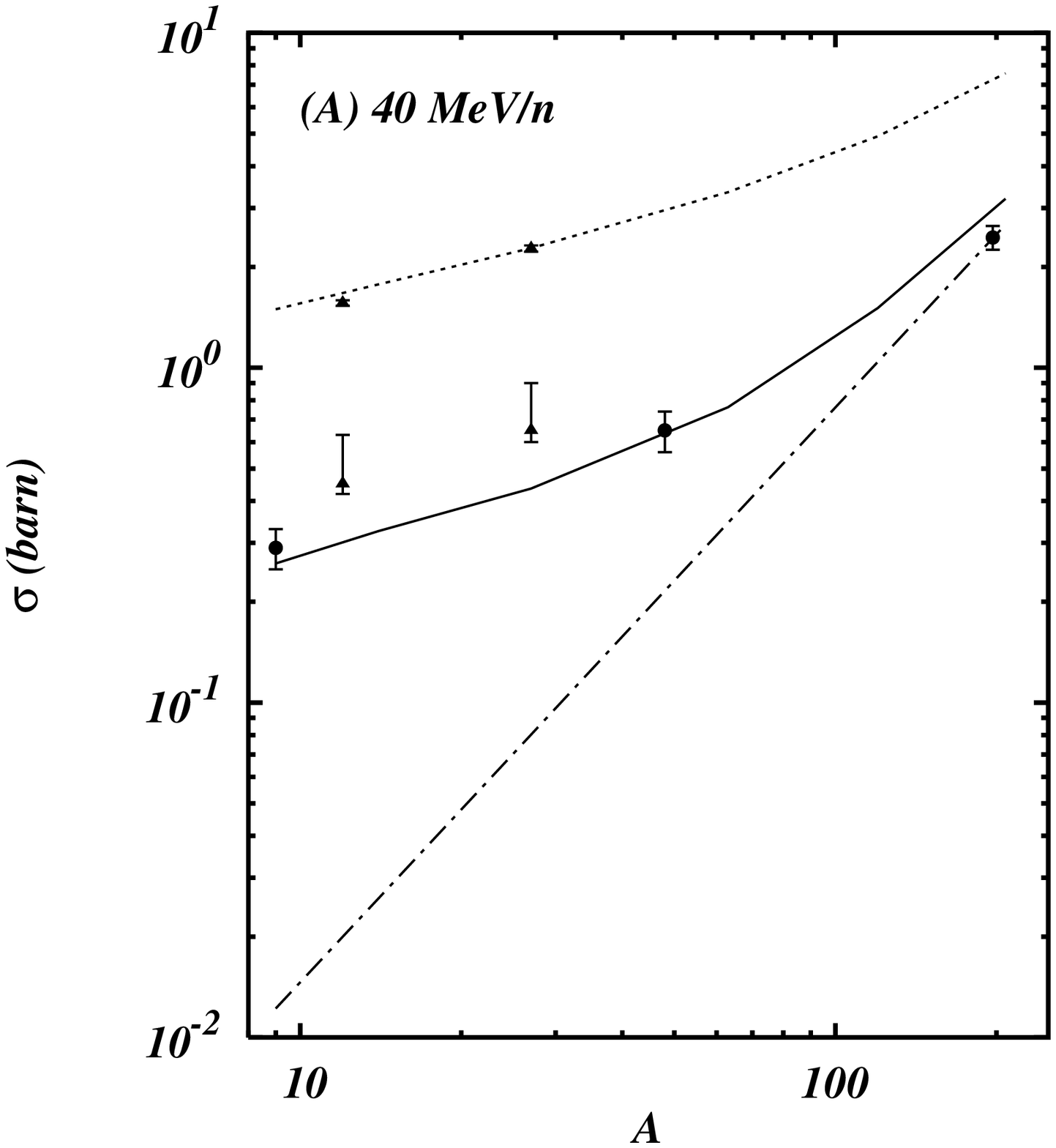}~~~~
\ForceWidth{0.44\hsize}
\tBoxedEPSF{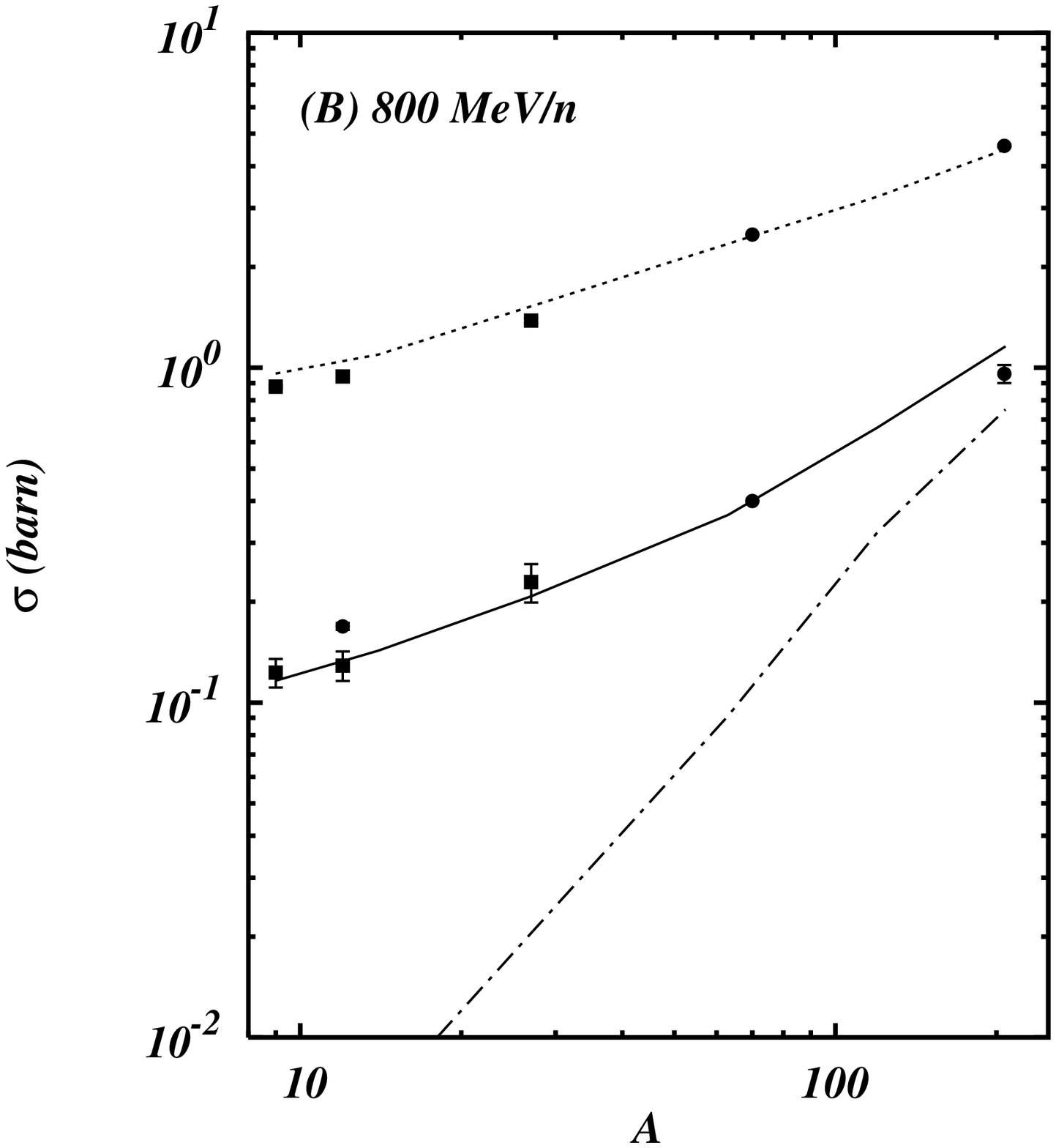}
\end{center}
\caption{Comparison of the neutron-removal cross section (solid line)
and the total interaction cross section (dashed line) with
experiments. We give the results again for an energy of 40 MeV/$n$
(A) and 800 MeV/$n$ (B). Also shown are the result for the Coulomb
breakup, which were added to our results.  The experimental results
are from
\protect\cite{anne94,fukuda91,tanihata88,tanihata85,kobayashi90}.
Please note that the results from \protect\cite{fukuda91} (triangles
in (A)) are for an energy of 33 MeV/$n$; the deviation from the
calculated results is therefore partly due to this.}
\label{Fig:exp}
\end{figure}

In Fig.~\ref{Fig:exp} we compare our results with experiments.  In
one kind of experiment, the core fragment is detected in the
reaction, so the processes that contribute are diffraction and
neutron absorption.  The solid lines show the theoretical cross
sections to which we have added Coulomb excitation cross sections
(dotted-dashed line) obtained in ref.~\cite{esbensen96b} from the
same single-particle model.  The theory agrees well with the data,
taken from refs
\cite{anne94,fukuda91,tanihata88,tanihata85,kobayashi90}.  The two
triangular data points in (A), which are somewhat high, are the
results of \cite{fukuda91} and are measured at an energy of 33
MeV/$n$. The squares for the one-neutron removal at 800 MeV/$n$ are
calculated by taken the difference of the interaction cross section
between $^{11}$Be and $^{10}$B in \cite{tanihata85}.  The agreement
at low energy is only possible because of the inclusion of the real
part of the potential, which enhances the diffraction component.  The
dotted line shows the calculated total interaction cross sections.
These are also in good agreement with experiment
\cite{anne94,fukuda91,tanihata88,tanihata85,kobayashi90}, as might be
expected for a quantity that is determined mainly by the geometric
size.
%
%

\begin{figure}[t]
\begin{center}
~\ForceWidth{0.50\hsize}
\tBoxedEPSF{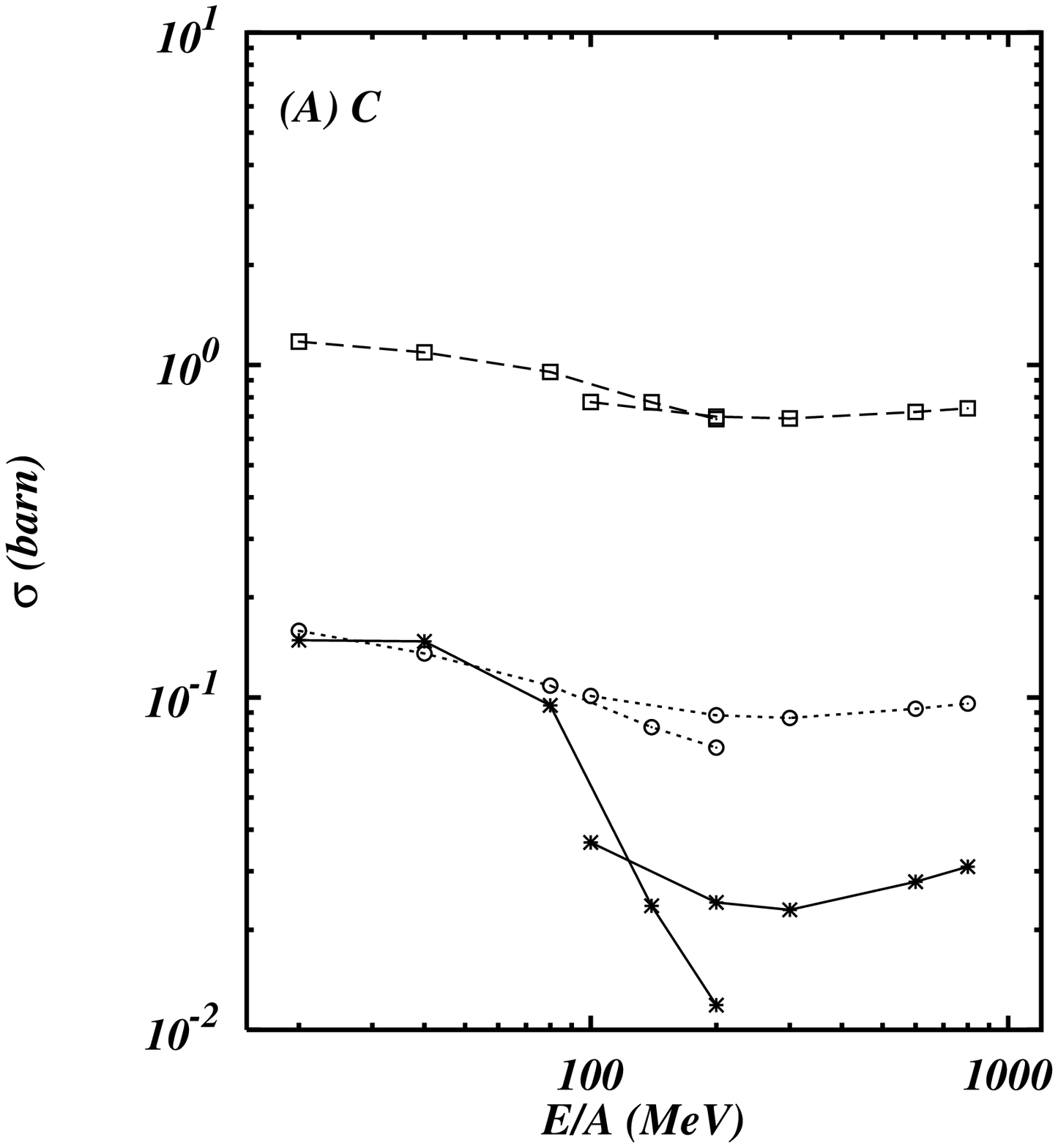}~
\end{center}
\caption{The dependence of the total cross sections on the projectile
energy $E$ is shown for a $C$ target. The notation is the same as in
Fig.~\protect\ref{Fig:adep}. The curves at lower energies are
calculated with the optical potential, the curves at higher energies
using the target density and free NN cross sections.}
\label{Fig:edep}
\end{figure}

In Fig.~\ref{Fig:edep} we show the dependence on the projectile
energy.  We have calculated the cross section using the optical
potential between 20 and 200~MeV/$n$ and using free NN cross sections
between 100 and 800~MeV/$n$. All cross section seem to have a smooth
transition between the low energy and high energy model between 100
and 200 MeV.
%
%

The energy dependence clearly illustrates the deviation of the cross
section from a black disk. The stronger dependence for a $C$ target
agrees with the expectation, as it is surface-dominated and therefore
less a black disk than heavier targets (not shown). We also see a
stronger dependence on the energy for those processes where a neutron
is involved, again in agreement with our expectation. At high
energies the total cross section follows more or less the energy
dependence of the NN cross section.  At low energies we have a rapid
decrease of the diffraction cross section at 100~MeV/$n$. This is due
to the decrease of the real part of the optical potential in this
region. As already discussed with Fig.~\ref{Fig:adep}, the cross
sections for diffraction and neutron-stripping are almost the same at
energies below 100~MeV/$n$ in agreement with other calculations
\cite{yabana92,anne93}.

\section{Differential cross sections for $^{11}\text{B}\lowercase{e}$}
\label{Sec:Be11}

We now discuss the differential cross sections for the momentum
distributions in the final state.  The neutron-stripping cross
section has often been calculated in an approximation called the
``transparent limit''.  There one neglects the effect of the
interaction between the observed fragment and the target nucleus and
uses the expression,
\begin{eqnarray}
\frac{d\sigma_{\text{n-str.,transp.}}}{d^3\vec{k}_c} 
&=& \frac{1}{(2\pi)^3} \frac{1}{2 L_0+1} \sum_{M_0}
\int d^2\vec{b}_n  
\left[1- \left|S_n \right|^2 \right] \times\nonumber\\
&&\left| \int d^3\vec{r} e^{-i \vec{k}_c\cdot\vec{r}} 
\phi_{0,M_0}(\vec{r}) 
\right|^2 \ ,
\end{eqnarray}
which follows from eq.~(\ref{Eq:strip}) simply by setting $S_c=1$.
With this approximation one gets a simple interpretation of the
momentum distribution as the Fourier transform of the wave function
of the ground state. The use of this approximation was questioned
recently \cite{esbensen96a}. First of all, the total cross section is
much too large.  Moreover, the additional factor of $S_c$ in
eq.~(\ref{Eq:strip}) weights the amplitudes more heavily where the
neutron is far from the core, and its momentum is lower.  This leads
to a narrower momentum distribution in the full theory.  Another
objection to the identification of final state momentum distributions
with the Fourier transform of the projectile wave function is that
the diffractive component does not behave this way at all, since the
final state is not a plane wave.

In the comparison with experimental data, we consider only the
longitudinal differential cross section. The longitudinal
distribution are much easier to calculate and to interpret. The
profile functions do not introduce any longitudinal momentum in the
system, so one is rather insensible to the details of the
potentials. When one looks at transverse distributions, there is not
only a dependence on the shape of the profile functions, but
higher-order scattering effects can also be significant. For the
kinematic regime we are treating here, elastic scattering affects the
transverse momentum much more strongly than the longitudinal
momentum. Furthermore we get a simple equation for the diffraction
only for the relative momentum. Relating this to the transverse
momentum distribution of either of the fragments is not trivial, but
requires a complete independent calculations, as different impact
parameter interfere coherently in this case.

\begin{figure}[t]
\begin{center}
\ForceWidth{0.50\hsize}
\BoxedEPSF{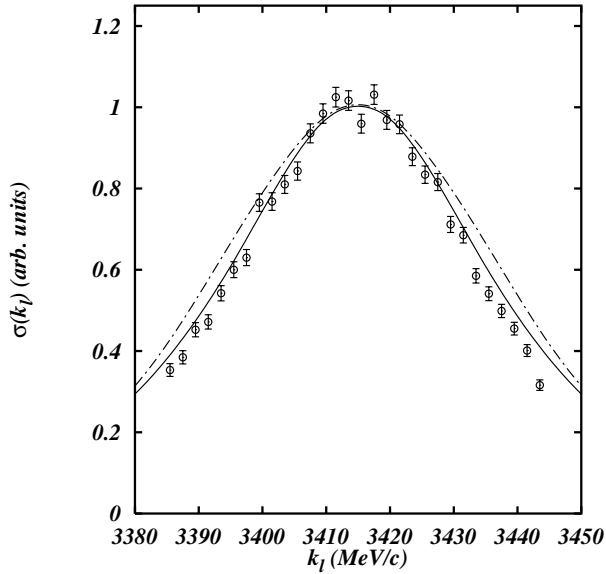}
\end{center}
\caption{Longitudinal momentum distribution of $^{10}$Be fragments,
produced in the breakup of $^{11}$Be at 66 MeV/$n$ on a $^9$Be
target.  The experimental results are from
\protect\cite{kelley95}. The full line is the result of the full
calculation for diffraction and stripping.  The dotted line is the
result of stripping in the transparent limit.}
\label{Fig:kelley}
\end{figure}

Let us now examine the longitudinal momentum distribution of the
$^{10}$Be fragment.  Fig.~\ref{Fig:kelley} shows the comparison of
theory with a measurement on a $^{9}$Be target at a beam energy of
66~MeV/$n$ \cite{kelley95}. The calculation of the solid curve
includes both diffraction and neutron absorption.  The two components
are of similar size and shape.  We show also the result of the
transparent limit as the dotted-dashed curve.  We find, in agreement
with ref.~\cite{esbensen96a}, that the full calculation has a
narrower momentum distribution than the transparent limit.
%
%

We have fitted the momentum distribution to a Lorentzian and also to
a sum of two Gaussians. From the fit to a Lorentzian we get a FWHM
$\Gamma$ of $41.4$ MeV/c for the full calculation and of $44.5$ MeV/c
for the transparent limit (always in the rest frame of
$^{11}$Be). The results of the full calculation are in good agreement
with the experimental result of $41.6\pm2.1$.  The individual
contributions have a width of $\Gamma=41.2$ for diffraction and of
$41.6$ for stripping.

Fitting to a sum of two Gaussians we get a good description of our
results.  The width $\sigma_0$ of the narrow component is $22.2$
MeV/c for the full calculation ($19.6$ MeV/c for diffraction and
$25.2$ for stripping) and $28.7$ MeV/c for the transparent limit.Our
fit to the data gives $\sigma_0=22.1$ MeV/c, again in good agreement
with the full calculation but not with the transparent limit.

\section{Differential cross sections for $^{8}$B}
\label{Sec:B8}

The breakup of $^{8}$B was measured in ref.~\cite{schwab95} at
1471~MeV/$n$ for different targets and interpreted using the
transparent limit of the Serber model.  The observed width was found
to be a factor of 2 smaller than the theory, showing that the
transparent limit is a poor approximation in this case. The
discrepancy is worse than for $^{11}$Be because of the $p$-wave
character of the halo nucleon. The $M=0$ state has the widest
distribution in the longitudinal direction because of its longitudinal
node, but it is suppressed in the full treatment because the proton
and the core are aligned along the beam axis. We calculate the
momentum distribution for a $^{12}$C target as total cross section are
given for this case. For the free NN cross section we use
$\sigma_{nn}=47$~mb and $\sigma_{np}=44$~mb.

\begin{figure}[t]
\begin{center}
\ForceWidth{0.50\hsize}
\BoxedEPSF{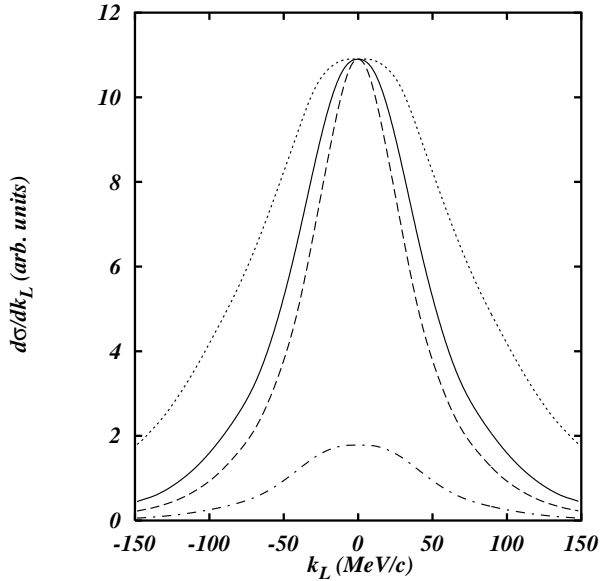}
\end{center}
\caption{Longitudinal momentum distribution of $^{7}$Be fragments,
produced in the breakup of $^{8}$B at 1470 MeV/$n$ on a $^{12}$C
target.  The full line is the result of the full calculation for
diffraction and stripping, the dash dotted line the contribution to
this from diffraction.  The dashed line is the contribution of the
$M\not=0$ states to the total result. The dotted line is the result
for stripping in the transparent limit.}
\label{Fig:schwab}
\end{figure}

The results for our calculation with a realistic profile function are
shown in Fig.~\ref{Fig:schwab}. The dotted line shows the result for
the transparent limit (only stripping), the solid line the result of
the full calculation. We show two contributions to the full
calculation: The dash-dotted line is the result for diffraction,
which contributes to the order of 20\%, the dashed line is the
contribution of the $M\not=0$ substates of the $p$ wave. As one can
see, the narrow width is mainly due to these states.
%
%

Fitting the curves with a Lorentzian, we get a FWHM $\Gamma$ of
88~MeV/c for the full calculation (93~MeV/c for diffraction
alone) and of 151~MeV/c for the transparent limit. Our results are
in agreement with the experimental result of $81\pm6$ MeV/c.

We get the total interaction cross section $\sigma_I({}^8\text{B})$ as
838~mb and the proton-removal cross section
$\sigma({}^8\text{B},{}^7\text{Be})$ as 69~mb in fair agreement with
the experimental results ($809\pm11$mb and $94\pm4$mb,
respectively). The discrepancy especially with the proton-removal
cross section can only partly be attributed to our folding model,
which is not a very good approximation for the core. A more detailed
model \cite{esbensen96b}, which uses the core wave functions, gets a
larger cross section but is still smaller than the experimental
result.

Comparing with the results at 790~MeV/$n$ and also for a $^{12}$C
target \cite{tanihata85,kobayashi96} we get a total interaction cross
section of 796~mb, which compares well with the recent experimental
value of $798\pm6$~mb.  The proton-removal cross section can be
calculated as the difference between the total cross sections of
$^{8}$B and $^{7}$Be. We get
$\sigma_I({}^8\text{B})-\sigma_I({}^7\text{Be})=60\pm11$~mb compared
to our result of $\sigma({}^8\text{B},{}^7\text{Be})=66$~mb.

The discrepancy between the 790 and 1471~MeV/$n$ data is
disturbing. As the difference between the free NN cross sections in
both cases is only of the order of 10\%, one would expect to get
similar results in both cases.  The proton-removal cross section is
only weakly sensible to the core interaction, but is more sensible to
the parameters of the potential, especially its radius. Using, for
example, a potential with $R=$2.678~fm at 1471~MeV/$n$ we get total
cross section of $\sigma_I({}^8\text{B})=842$~mb and
$\sigma({}^8\text{B},{}^7\text{Be})=74$~mb. The proton removal cross
section could therefore be used to calibrate the radius of the
potential. But both measurements would give different results for
this. It is therefore important to resolve this discrepancy, for
example, by a direct measurement of $\sigma_{1p}$ at 790~MeV/$n$. The
determination of the size of the core potential is crucial for the
prediction of the $S$-factor for the $p+{}^7Be \rightarrow {}^8{B}$
radiative capture process, see, e.g, ref.~\cite{brown96}.

\section{Summary and Conclusions}
\label{Sec:Summary}

The experimental results for the nuclear induced breakup of $^{11}$Be
on light targets, namely the total cross section and the longitudinal
momentum distribution of $^{10}$Be fragments, can be reproduced in
the eikonal model using a single-particle description of the
halo-nucleon and using realistic potentials to generate the profile
functions.  The real part of the optical potential is clearly needed
at lower beam energies (say below 100 MeV/$n$) in order to reproduce
the measured one-neutron removal cross sections.  At higher energies
the interaction deviates from the black disk model due to the
smallness of the nucleon-nucleon interaction.

In order to get a good agreement with measured longitudinal momentum
distributions of $^{10}$Be fragments, one needs to use the full
expression for the differential cross section. Calculations based on
the transparent limit of the stripping process, on the other hand,
produce distributions that are to wide (resulting in a prediction of
a wider halo radius if fit to the data). At lower beam energies,
diffractive and stripping processes produce longitudinal momentum
distributions that are similar in magnitude and width.  At higher
beam energies, stripping starts to dominate the one-neutron removal.

We have calculated also total cross sections as well as longitudinal
momentum distribution for the breakup of $^{8}$B. As the ground state
of the proton is a $p$ state, the discrepancy between transparent
limit and full calculation is much more drastic in this case.  The
width of the full calculation is in good agreement with the
experiments; the total cross sections are in fair agreement, with a
discrepancy between the data of the two experiments.

At the moment our calculation are restricted to longitudinal momentum
distributions. These are less sensible to the details of the
interaction and more sensible to the structure of the halo and are
therefore more easily interpreted. In the future we want to extend
our calculations also to the transverse momentum distribution. In
this case we have to include also the Coulomb potential in our
calculation, especially in the case of $^{8}$B.  In the case of the
diffraction we get a simple equation only for the relative momentum
distribution between nucleon and core, which normally is not measured
in experiments. But as measurements of the transverse momentum
distribution have been done and more will be done in the future, it
is interesting to see, whether a single particle model can describe
their outcome.

\acknowledgments
\label{Sec:acknow}
This work was supported in part by the Swiss National Science
Foundation (SNF) and the ``Freiwillige Akademische Gesellschaft''
(FAG) of the University of Basel, and by the U.S. Department of
Energy, Nuclear Physics Division, under Contracts DE-FG-06-90ER-40561
and W-31-109-ENG-38.


\end{document}